\documentclass[12pt]{article}
\usepackage{etoolbox}
\usepackage{graphicx}
\usepackage{pgfplots}
\usepackage{booktabs}
\usepackage{amsmath}
\usepackage{algorithm}
\usepackage{algorithmic}
\usepackage{amssymb}
\usepackage{amsfonts}
\usepackage{latexsym}
\usepackage{color}
\usepackage{subfigure}
\usepackage{latexsym}
\usepackage{graphicx,epstopdf}
\usepackage{ulem}

\usepackage{authblk}
\usepackage{caption}
\usepackage[utf8]{inputenc}
\usepackage{esint}
\usepackage{tikz}
\usepackage{amsthm}
\usetikzlibrary{arrows,decorations.markings,calc}
\usetikzlibrary{decorations.pathmorphing,patterns}
\usepackage{makeidx}
\makeindex
\catcode `\@=11 \@addtoreset{equation}{section}

\catcode `\@=12

\newcommand{\be}{\begin{equation}}
	\newcommand{\en}{\end{equation}}
\newcommand{\bea}{\begin{eqnarray}}
	\newcommand{\ena}{\end{eqnarray}}
\newcommand{\beano}{\begin{eqnarray*}}
	\newcommand{\enano}{\end{eqnarray*}}
\newcommand{\bee}{\begin{enumerate}}
	\newcommand{\ene}{\end{enumerate}}

\newcommand{\R}{{\cal R}}

\newcommand{\Hil}{{\cal H}}
\newcommand{\I}{{\cal I}}

\newcommand{\Lc}{{\cal L}}
\newcommand{\D}{{\cal D}}

\newcommand{\K}{{\mathcal K}}
\newcommand{\V}{{\mathcal V}}

\newcommand{\Sc}{{\cal S}}

\newcommand{\A}{{\cal A}}
\newcommand{\1}{1 \!\! 1}
\newcommand{\ha}{\hat{a}}
\newcommand{\had}{\hat{a}^\dagger}

\catcode `\@=11 \@addtoreset{equation}{section}
\catcode `\@=12

\begin{document}
\vspace*{-2cm}
\vspace{1cm}
\noindent This is the AAM reviewed version of the article published at Philosophical Transactions A. \\
doi: 10.1098/rsta.2024-0619
\thispagestyle{empty}

\begin{center} {\Large \bf Modeling Epidemics with Memory Effects: an
		Open Quantum System Approach}   \vspace{1cm}\\

{\large F. Bagarello}\\
  Dipartimento di Ingegneria,
Universit\`a di Palermo, 90128  Palermo, Italy\\
and I.N.F.N., Sezione di Catania, 95123 Catania, Italy\\
e-mail: fabio.bagarello@unipa.it\\

\vspace{2mm}

{\large F. Gargano}\\
Dipartimento di Ingegneria,
Universit\`a di Palermo, 90128  Palermo, Italy\\
e-mail: francesco.gargano@unipa.it\\

\vspace{2mm}

{\large P. Khrennikova}\\
Faculty of Behavioural, Management, and Social Sciences, University of Twente, Netherlands\\
e-mail: p.khrennikova@utwente.nl\\

\end{center}

\vspace{0.5cm}
\begin{abstract}
	In this work, we introduce a quantum-inspired epidemic model to study the dynamics of an infectious disease in a population divided into  compartments. By treating the healthy population as a large reservoir, we construct a framework based on open quantum systems and a Hilbert space formalism to model the spread of the infection. This approach allows for a mathematical framework that captures both Markovian and semi-Markovian dynamics in the evolution equations.
	Through numerical experiments, we examine the impact of varying memory parameters on the epidemic evolution, focusing in particular on the conditions under which the model remains physically admissible.
\end{abstract}

%{\bf PACS Numbers}:  .......

%\vfill

%\pagenumbering{roman}

\newpage
\section{Introduction}
In recent years, there has been a notable increase in the application of quantum methods beyond the traditional microscopic realm. 
This growing interest has led to a rich literature, where quantum tools--such as Hilbert spaces, operator theory, and functional analysis-- have been employed in a variety of unexpected fields, including decision-making, psychology, economics, finance, biology, and others. 
\noindent We only cite here some monographs: \cite{bagbook1}-\cite{havkhrebook}, where many other references can be found.
Few years ago, in \cite{bff}, shortly after the appearance of COVID-19, the authors have applied a full Hamiltonian strategy, based on the general theory of open quantum systems, to describe the growth of the epidemics, in its first wave. That paper had some limitation mainly due to the Hamiltonian approach proposed there, which made it unavoidable to include processes which are not really possible (since, e.g., infected people can die, the authors had to consider a small possibility that dead people can, in principle, become infected\footnote{It should be stressed that this transition was not really observed because of the initial conditions adopted in \cite{bff}.}.) 
Also, all interactions in \cite{bff} were assumed to be instantaneous, which is a strong assumption and not entirely realistic. In this paper, we aim to address both of these limitations by using the Gorini-Kossakowski-Sudarshan-Lindblad (GKSL) evolution equation, which efficiently describes quantities that exhibit a suitable asymptotic behavior and taking into account of memory effects within the system, according to a semi-Markovian perspective.
Our framework contributes to and advances the broader body of research that applies the Markovian (memoryless) version of the GKSL model to various social, political, biological, and cognitive systems. For instance, \cite{Elections2014} and \cite{Elections2016} applied the open systems approach to capture the stabilization of voters' preferences in interaction with the mass-media environment. The open systems model was also applied to experimental data on preference evolution by \cite{BuseGKSL}, capturing the oscillation of preference formation driven by the indeterminate internal state of the decision maker and the informational environment of outside observers. In the biological domain, \cite{Biosystems2021} modeled various biological processes as interactions with an information environment.
%perhaps, also your other papers on political alliances etc were you used open systems? 

The remainder of the paper is organized as follows: After this introduction, Section \ref{sect2} provides a general overview of the mathematical and physical framework of the composite system we aim to describe. In Section \ref{sect3}, we present the detailed description of this system, including the semi-Markovian time evolution. Section \ref{sect4} focuses on a specific application of the framework introduced in Section \ref{sect3}, namely, the analysis of a particular epidemic model. The numerical analysis is presented in Section \ref{sect5}, and Section \ref{sect6} concludes the paper with a summary of our findings, future research directions, and a possible alternative interpretation of the model from Section \ref{sect4}, more focused on decision-making.

\section{The Open Quantum System approximation}\label{sect2}
We assume a epidemic model for the evolution of a population divided primarily into four distinct compartments: {Healthy} (\(\mathcal{H}\)), {Infected} (\(\mathcal{I}\)), {Recovered} (\(\mathcal{R}\)), and {Dead} (\(\mathcal{D}\)). We consider the natural assumption that, during the time evolution of the system, healthy individuals significantly outnumber the other compartments. Under this assumption, we can approximate the entire system as being represented by a large reservoir of individuals, whose dynamics can be considered relatively stable over the timescale of interest, and a small system that is heavily influenced by the presence of healthy individuals due to the spread of the epidemic.  
Following the approach that two of us employed some years ago \cite{bff}, we seek to describe the dynamics of the epidemic using the framework of an open quantum system, but in a slightly different way, as we will explain. By considering \(\mathcal{H}\) as a large reservoir, we can focus on the evolution of the smaller system \(\mathcal{S}_{\mathcal{P}}\), made by \(\mathcal{I}\), \(\mathcal{R}\), and \(\mathcal{D}\), without explicitly modeling the entire population. This approximation is particularly useful because it simplifies the mathematical analysis while capturing the essential features of the epidemic dynamics. In other words, we will consider here $\mathcal{H}$ as a sort of infinite source of individuals, which is not significantly affected when some of them become infected.

Our main goal is to determine the densities of individuals in the compartments of \(\mathcal{S}_{\mathcal{P}}\). We model interactions between compartments, assuming that the infectious process increasing $\mathcal{I}$ is instantaneous and {\it time-local}, meaning the probability of a healthy individual becoming infected depends only on the current state, with no influence from past states.
Conversely, transitions from the infected state to recovered or dead compartments are modeled as semi-Markovian processes. In these processes, the probability of transitioning to a new state depends not only on the current state but also on what happens after entering that state. This reflects the realistic progression of many diseases, where infection duration influences recovery or death likelihood. For example, recovery probabilities for seasonal flu might follow a normal distribution, as would the death process. We should stress that introducing a {\it time for recovering} or a {\it time for dying} is, in our view, an improvement over the model in \cite{bff}, where all transitions were instantaneous.

\begin{figure}[!h]\centering
\tikzset{every picture/.style={line width=0.75pt}} %set default line width to 0.75pt        

\begin{tikzpicture}[scale=0.85,x=0.75pt,y=0.75pt,yscale=-1,xscale=1]
	
	%Rounded Rect [id:dp9717346797382345] 
	\draw   (106,76.84) .. controls (106,72.51) and (109.51,69) .. (113.84,69) -- (182.35,69) .. controls (186.68,69) and (190.19,72.51) .. (190.19,76.84) -- (190.19,100.34) .. controls (190.19,104.67) and (186.68,108.18) .. (182.35,108.18) -- (113.84,108.18) .. controls (109.51,108.18) and (106,104.67) .. (106,100.34) -- cycle ;
	%Rounded Rect [id:dp3663548864084345] 
	\draw   (234,58) .. controls (234,53.58) and (237.58,50) .. (242,50) -- (296,50) .. controls (300.42,50) and (304,53.58) .. (304,58) -- (304,82) .. controls (304,86.42) and (300.42,90) .. (296,90) -- (242,90) .. controls (237.58,90) and (234,86.42) .. (234,82) -- cycle ;
	%Rounded Rect [id:dp5573832000264738] 
	\draw   (348,79) .. controls (348,74.58) and (351.58,71) .. (356,71) -- (410,71) .. controls (414.42,71) and (418,74.58) .. (418,79) -- (418,103) .. controls (418,107.42) and (414.42,111) .. (410,111) -- (356,111) .. controls (351.58,111) and (348,107.42) .. (348,103) -- cycle ;
	%Rounded Rect [id:dp06170402313450629] 
	\draw  [line width=2.25]  (180.44,178.9) .. controls (180.44,171.99) and (186.04,166.39) .. (192.95,166.39) -- (356.93,166.39) .. controls (363.84,166.39) and (369.44,171.99) .. (369.44,178.9) -- (369.44,216.43) .. controls (369.44,223.34) and (363.84,228.94) .. (356.93,228.94) -- (192.95,228.94) .. controls (186.04,228.94) and (180.44,223.34) .. (180.44,216.43) -- cycle ;
	%Straight Lines [id:da803115090251936] 
	\draw    (270.15,162.44) -- (269.23,102.18) ;
	\draw [shift={(269.19,99.18)}, rotate = 89.13] [fill={rgb, 255:red, 0; green, 0; blue, 0 }  ][line width=0.08]  [draw opacity=0] (8.93,-4.29) -- (0,0) -- (8.93,4.29) -- cycle    ;
	%Straight Lines [id:da8896730280486638] 
	\draw    (197.97,85.05) -- (232.19,71.18) ;
	\draw [shift={(195.19,86.18)}, rotate = 337.93] [fill={rgb, 255:red, 0; green, 0; blue, 0 }  ][line width=0.08]  [draw opacity=0] (8.93,-4.29) -- (0,0) -- (8.93,4.29) -- cycle    ;
	%Straight Lines [id:da06509116242643165] 
	\draw    (309.19,72.18) -- (340.5,87.84) ;
	\draw [shift={(343.19,89.18)}, rotate = 206.57] [fill={rgb, 255:red, 0; green, 0; blue, 0 }  ][line width=0.08]  [draw opacity=0] (8.93,-4.29) -- (0,0) -- (8.93,4.29) -- cycle    ;
	%Shape: Ellipse [id:dp3527669502197601] 
	\draw  [line width=1.5]  (68,86.94) .. controls (68,54.35) and (160.27,27.94) .. (274.09,27.94) .. controls (387.92,27.94) and (480.19,54.35) .. (480.19,86.94) .. controls (480.19,119.52) and (387.92,145.94) .. (274.09,145.94) .. controls (160.27,145.94) and (68,119.52) .. (68,86.94) -- cycle ;
	%Shape: Ellipse [id:dp6095299529501821] 
	\draw  [line width=2.25]  (16.44,123.91) .. controls (16.44,57.35) and (133.97,3.39) .. (278.94,3.39) .. controls (423.92,3.39) and (541.44,57.35) .. (541.44,123.91) .. controls (541.44,190.48) and (423.92,244.44) .. (278.94,244.44) .. controls (133.97,244.44) and (16.44,190.48) .. (16.44,123.91) -- cycle ;
	
	% Text Node
	\draw (106,79.84) node [anchor=north west][inner sep=0.75pt]   [align=left] {\small Recovered};
	% Text Node
	\draw (235,63) node [anchor=north west][inner sep=0.75pt]   [align=left] {\small Infected};
	% Text Node
	\draw (363,82) node [anchor=north west][inner sep=0.75pt]   [align=left] {Dead};
	% Text Node
	\draw (243,202) node [anchor=north west][inner sep=0.75pt]   [align=left] {Healthy};
	% Text Node
	\draw (434,71) node [anchor=north west][inner sep=0.75pt]  [font=\large] [align=left] {$\displaystyle \mathcal{S}_{P}$};
	% Text Node
	\draw (310,185) node [anchor=north west][inner sep=0.75pt]  [font=\large] [align=left] {$\displaystyle \mathcal{H}$};
	% Text Node
	\draw (419,165) node [anchor=north west][inner sep=0.75pt]  [font=\LARGE] [align=left] {$\displaystyle \mathcal{S}$};
\end{tikzpicture}
\end{figure}
%\subsection{Hilbert Space Construction for Epidemic Dynamics}
To formalize the above assumptions, 
we consider a composite Hilbert space representing the states of the population divided into three categories: infected ($\I$), recovered ($\R$), and dead ($\D$). Each of these populations is associated with a Hilbert space of dimension 2, where the basis states are $\lvert 0 \rangle$ and $\lvert 1 \rangle$, representing the absence or presence of individuals in each category.
Let $\mathcal{H}_{\I}$, $\mathcal{H}_{\R}$, and $\mathcal{H}_{\D}$ denote the Hilbert spaces for infected, recovered, and dead populations, respectively. The full Hilbert space of the system is then given by:
\be
\mathcal{H}_{full} = \mathcal{H}_{\I} \otimes \mathcal{H}_{\R} \otimes \mathcal{H}_{\D},
\en
which means that $\Hil_{full}$ is the tensor product of the three compartments spaces,
$\mathcal{H}_{\I}$, $\mathcal{H}_{\R}$, and $\mathcal{H}_{\D}$, corresponding to each sub-population. We observe that the health population is not included here, in view of what we have discussed before. The basis states of $\mathcal{H}_{full}$ are defined by tensor products of individual states, such as  $\lvert \I \rangle \otimes \lvert \R \rangle \otimes \lvert \D \rangle$, where each component corresponds to the specific status (0 or 1) of the population in each compartment. For instance the vector $\lvert 0 \rangle \otimes \lvert 0 \rangle \otimes \lvert 0 \rangle$ describes a situation in which there are no infected, recovered and dead people: this is a natural initial state of the time evolution, where all the population is healthy, so that $\Sc_p$ is completely empty. Of course, within our scheme the dimensionality of the total Hilbert space is $8$.

As in similar approaches on several dynamical systems considered using operator techniques, \cite{bagbook1,bagbook2,fffbook}, we can attach some ladder operators to each compartment of the system: \(\hat{a}_{\I}\) and \(\hat{a}_{\I}^\dagger\) for the infected,  \(\hat{a}_{\R}\) and \(\hat{a}_{\R}^\dagger\)  for the recovered, \(\hat{a}_{\D}\) and \(\hat{a}_{\D}^\dagger\) for the dead. These are the typical (fermionic) annihilation and creation operators, which are labeled according the compartments, but are essentially constructed out of the following  $2\times 2$ matrices: 
\be
\ha_{\I,\R,\D} = 
\begin{pmatrix}
0 & 1 \\
0 & 0
\end{pmatrix}, \qquad \had_{\I,\R,\D} = 
\begin{pmatrix}
0 & 0 \\
1 & 0
\end{pmatrix}.
\en
Each of these operators has a natural extension to $\Hil_{full}$. For instance, $\hat a_\I\rightarrow\hat a_\I\otimes\mathbb{I}_{\R}\otimes\mathbb{I}_{\D}$. However, quite often, when not needed, we will simply adopt the symbol $\hat a_\I$.
The above operators satisfy the  rules
$
\{\hat{a}_{n}, \hat{a}_{m}^\dagger\} = \delta_{n,m} \1,
$
for all $n, m = \I, \R, \D$. Here, $\{x, y\} := xy + yx$, and all other anti-commutators are  zero.  {The use of fermionic modes is advantageous from a modeling perspective because, as explained below, they ensure that the expectation value of the number operator	
for each compartment remains within the interval $[0,1]$, which is what should happen for {\it densities of populations}. This follows from the fermionic rules $\hat a_n^2=(\hat a_n^\dagger)^2=0$ for all $n=\I,\R,\D$.}
To obtain the expectation value of each population, we use the trace operation with the {number} operator $N_{\I,\R,\D}=\had_{\I,\R,\D}\ha_{\I,\R,\D}=\lvert 1 \rangle \langle 1 \rvert_{\I,\R,\D}$ in each subspace\footnote{Indeed a raising operator can be written, in a bra-ket language, as $\hat a^\dagger=\lvert 1 \rangle \langle 0 \rvert$, so that its adjoint is $\hat a=\lvert 0 \rangle \langle 1 \rvert$. Hence, using the fact that $\lvert 0 \rangle$ is normalized, $\hat N=\hat a^\dagger \hat a=\lvert 1 \rangle \langle 1 \rvert$.}:
\be\label{ExValues}
\langle \I \rangle = \text{Tr}(\rho \, (\lvert 1 \rangle \langle 1 \rvert_\I \otimes \mathbb{I}_{\R} \otimes \mathbb{I}_{\D})),\,
\langle \R \rangle = \text{Tr}(\rho \, (\mathbb{I}_{\I} \otimes \lvert 1 \rangle \langle 1 \rvert_\R \otimes \mathbb{I}_{\D})),\,
\langle \D \rangle = \text{Tr}(\rho \, (\mathbb{I}_{\I} \otimes \mathbb{I}_{\R} \otimes \lvert 1 \rangle \langle 1 \rvert_\D)).
\en
These expressions give the expectation values for the (densities of) infected, recovered, and dead populations by taking the trace over the density matrix with the respective number  operator in each subspace.
% can we perhaps add a social interpretation of densities?  Such as the average frequency of   infected/recovered/ dead, or as part of a probabilistic prognosis, the subjective probability, the chance to recover/die conditioned on the infected state and memory effect. 

\section{A general hybrid model with memory effect}\label{sect3}

{In this section we will discuss a general hybrid model in presence of memory effects based on a Lindblad-like superoperator. This model will be specialized %adapted?
to a concrete situation in Section \ref{sect4}. 

Let us consider a  finite set $\{|n\rangle\}$ of states forming an orthonormal basis of a finite-dimensional Hilbert space {\it attached to} the open quantum system we are interested to describe.}
We consider the hybrid time evolution of the reduced density operator, incorporating both time-local local and non-local time-delayed effects. The evolution equation we shall consider is given by
\be \label{eq1}
\frac{d}{dt} \rho(t) = \Lc(t)[\rho(t)] + \int_0^t d\tau \, \K(\tau)[\rho(t - \tau)],
\en
The superoperator \(\K(t)\), which accounts for memory effects, is defined as
\be \label{eq2}
\K(t)[\bullet] = -i [H(t),\bullet]+ \sum_{mn} k_{mn}(t) \left(|m\rangle\langle n|\bullet |n\rangle\langle m|  - \frac{1}{2}  \{ |n\rangle\langle n|, \bullet \}\right) ,
\en
with the Hamiltonian $\label{hamil}
H(t) = \sum_n \varepsilon_n(t) |n\rangle\langle n|.
$
{The set $\{|n\rangle\}$ is therefore an orthonormal basis of eigenvectors of the time-dependent Hamiltonian $H(t)$. Moreover, these vectors are also eigenstates of $H^{nc}(t)$, see below. We refer to (\ref{41}) for an explicit example of these eigenvectors useful for our application.}
The superoperator \(\Lc(t)\), which accounts only for time-local effects, is defined as
\be \label{eq3}
\Lc(t)[\bullet] = -i [H^{nc}(t),\bullet]+ \sum_{mn} k^{nc}_{mn}(t) \left(|m\rangle\langle n|\bullet |n\rangle\langle m|  - \frac{1}{2}  \{ |n\rangle\langle n|, \bullet \}\right),
\en
with the Hamiltonian
$ \label{hamilnc}
H^{nc}(t) = \sum_n \varepsilon^{nc}_n(t) |n\rangle\langle n|.
$
Here, the superscript \({}^{nc}\) denotes operators and functions associated solely with the time-local effects, reflecting the non-convolutional nature of \(\Lc\).
All the functions $k_{nm}^\sharp(t),\varepsilon_n^\sharp(t)$ are considered non negative for all time.
Note that, as already observed, we are assuming that both the Hamiltonians above are diagonal in the basis \(\{|n\rangle\}\), and \eqref{eq2} and \eqref{eq3} correspond to the standard Lindblad dissipators appearing in the GKSL equations, where the Lindblad operators are of the form $L_{m,n}=|m\rangle\langle n|$.
Given that we are using Lindblad operators, the preservation of trace and Hermiticity in the dynamical map \( \V(t) \) we will introduce later is inherently satisfied. However, it remains crucial to ensure the preservation of positivity in the evolution of the density operator. This is particularly needed in view of our interpretation, where the density operator is linked to concrete {\it densities of populations}.
While the Lindblad form with only the time-local part of the evolution (described by \(\Lc\)) preserves positivity (and complete positivity), the inclusion of non-local memory effects through the superoperator \(\K\) requires careful consideration (\cite{BreuerVacchini2008}). We will discuss more on this aspect later on.
\subsection{The equations of motion}
First, we note that for any operator $|n\rangle\langle m|$, we have
\be \label{eq7}
\A(t)\big[|n\rangle\langle m|\big] = \left(-z^{\sharp}_n(t) - \overline{z^{\sharp}_m}(t)\right) |n\rangle\langle m| + \delta_{n,m} \sum_{j}  k^{\sharp}_{jn}(t) |j\rangle\langle j|,
\en
where \( \A \) can represent either \( \Lc \) or \( \K \), and
$
z^{\sharp}_n(t) =\left( \frac{1}{2} \sum_j k_{jn}^{\sharp}(t)\right) + i \varepsilon^\sharp_n(t),
$
with the superscript \(\sharp\) indicating the time-local and non-local functions appearing in equations \eqref{eq2} and \eqref{eq3}.
Considering the action described by equation \eqref{eq7} and substituting in equation \eqref{eq1}, we can derive the governing equations for the elements  \( \rho_{nm}(t) \):
\bea
\frac{d}{dt} \rho_{nm}(t) &=& \left(-z_n^{nc}(t) - \overline{z_m^{nc}}(t)\right) \rho_{nm}(t) + \delta_{n,m} \sum_j  k^{nc}_{nj}(t) \rho_{jj}(t) \nonumber \\
&& - \int_0^t d\tau \, \left[ \left(z_n(\tau) + \overline{z_m}(\tau)\right) \rho_{nm}(t-\tau) - \delta_{n,m} \sum_j  k_{nj}(\tau) \rho_{jj}(t-\tau) \right].
\ena
Therefore, the special representation of $\Lc$ and $\K$ allows to write in a closed form the equation for the populations $\rho_{nn}$ in the following generic  set of master equations:
\bea \label{eqpopulations}
\frac{d}{dt} \rho_{nn}(t) = \sum_j \big[k^{nc}_{nj}(t) \rho_{jj}(t) -k_{jn}^{nc}(t) \rho_{nn}(t)\big]+ \int_0^t d\tau \, \sum_j \big[ k_{nj}(\tau) \rho_{jj}(t-\tau)- k_{jn}(\tau)  \rho_{nn}(t-\tau) \big],
\ena
whereas the coherences $\rho_{nm},\,n\neq m$ are only dependent on their history and satisfy the simpler equations
\bea \label{eqcoherences}
\frac{d}{dt} \rho_{nm}(t) &=& \left(-z_n^{nc}(t) - \overline{z_m^{nc}}(t)\right) \rho_{nm}(t) - \int_0^t d\tau \, \left[ \left(z_n(\tau) + \overline{z_m}(\tau)\right) \rho_{nm}(t-\tau)  \right].
\ena
The system of equations \eqref{eqpopulations} is particularly important because, if the initial condition for the density operator is such that only the diagonal elements \(\rho_{nn}(0)\) are non-zero, then only the evolution of the populations \(\rho_{nn}(t)\) is relevant.
In general the evolution must ensure that for any initial density matrix \(\rho(0)\), the evolved state \(\rho(t)\) remains a valid density matrix which is, in particular, positive definite. In the context of the equations derived here, this means that the populations \( \rho_{nn}(t) \) must remain non-negative at all times, and hence that the functions \(k^{\sharp}_{nm}(t)\)  should be carefully chosen to avoid introducing any terms that could lead to negative populations. 

\subsection{The condition for the positivity of the evolution}
We now want to derive some necessary and, under certain hypothesis, sufficient conditions for the positivity of the evolution.
Following and extending what done \cite{BreuerVacchini2008} for a fully semi-Markovian process (without the add of time local effects), from the derived equations for the populations and coherences, \eqref{eqpopulations} and \eqref{eqcoherences},
we can directly seek a dynamical map \( \V(t) \) such that
\be
\rho(t) = \V(t)[\rho(0)],
\en
where we write, in full generality, 
\be\label{dynmap}
\V(t)[\rho(0)] = \sum_{n \neq m} g_{nm}(t) |n\rangle\langle n| \rho(0) |m\rangle\langle m| + \sum_{nm} T_{nm}(t) |n\rangle\langle m| \rho(0) |m\rangle\langle n|,
\en
which expresses how each coherence only depends on its own evolution, whereas the populations $\rho_{nn}$ form a closed system. Given that \( \V(0) = \mathbb{I} \), the time-zero coefficients of this expansion must satisfy \( g_{nm}(0) = 1 \) and \( T_{nm}(0) = \delta_{nm} \) for each \( n, m \).

Substituting \( \V(t)[\rho(0)] \) into \eqref{eq1} and using \eqref{eq7}, we find that the \( g_{nm}(t) \) satisfy the same equations as in \eqref{eqcoherences} with the initial condition \( g_{nm}(0) = 1 \). To write the equation for \( T_{nm}(t) \), for a specific $m$, we start from the initial condition \( \rho_{mm}(0) = 1 \), and due to the orthogonality of the diagonal matrices \(\{|n\rangle\langle n|\}\), after differentiating, we have:
\be
\dot{\rho}(t) = \dot{\V}(t)[\rho(0)] = \sum_n \dot{T}_{nm}(t) |n\rangle\langle n|.
\en
Applying the action of \(\Lc\) and \(\K\) on \(\V(t)[\rho(0)]\), we finally get:
\bea
\hspace*{-0.5cm}\frac{d}{dt} T_{nm}(t) = \sum_j \left[k^{nc}_{nj}(t) T_{jm}(t) - k^{nc}_{jn}(t) T_{nm}(t)\right]+\int_0^t d\tau \sum_j \left[k_{nj}(\tau) T_{jm}(t - \tau) - k_{jn}(\tau) T_{nm}(t - \tau)\right]. \hspace*{0.1cm}\nonumber \\ \label{transProb}
\ena 
For each \( n \), the equation for \( T_{nn}(t) \) is the same as \eqref{eqpopulations} with the initial condition \( T_{nn}(0) = 1 \).
The above set of equations form a set of generalized master equations for the functions \( T_{mn}(t) \), which can now be interpreted as transition probabilities from state \( |n\rangle \) at $t=0$ to state \( |m\rangle \) over time \( t \), provided they are always non-negative. These equations account for both Markovian (memoryless) and non-Markovian effects, with the latter due to memory kernels representing delayed effects of transitions between states.
{
In particular, since $k^{nc}_{nj}(t)$ is the rate of the transitions from $|j\rangle$ to $|n\rangle$,  the time local term  $\sum_j k^{nc}_{nj}(t) \, T_{jm}(t)$
can be interpreted as the probability gained for transitioning into the state $ |n\rangle $ from intermediate states $|j\rangle$, starting from $ |m\rangle$ at $t=0$. Conversely, $-\sum_j k^{nc}_{jn}(t) \, T_{nm}(t)$ represents the probability which is lost due to the process of leaving $|n\rangle$ to go to other states $|j\rangle$ (again starting from $|m\rangle$ at $t=0$). A similar interpretation applies to the integral terms, with the key difference that those terms incorporate memory effects through the convolution over the past.}
Actually the system of equations \eqref{eqpopulations} and \eqref{transProb}, with the appropriate initial conditions,  are a generalization of the governing equations equations for a semi Markov process, see \cite{Feller1964, BreuerVacchini2008}, with the inclusion of a time local effect. 
{A key aspect of the dynamics governed by such a  process is the definition of the survival probability functions $ g_{n}(t)$, which define the probability that the system has not left state \( |n\rangle \) after a certain time $t$, starting from it at $t=0$, hence a non negative function.}  
{Extending the usual definition, see again \cite{BreuerVacchini2008}, and considering both local time and memory effects, we can here define the functions $ g_{n}(t) $ by means of the following system of integro-differential equations
\bea \label{eqgnn}
\frac{d}{dt} g_{n}(t) &=& -\sum_j k_{jn}^{nc}(t) g_{n}(t) - \int_0^t d\tau \, \sum_j k_{jn}(\tau) g_{n}(t-\tau),
\ena
with the initial condition $ g_{n}(0) = 1 $.
The above equations can be deduced from \eqref{transProb} considering that in the context of survival, we are interested only in the probability of not leaving state $|n\rangle$ without intermediate jumps, and hence considering the only contributions coming from the loss terms. Hence $-\sum_j k_{jn}^{nc}(t) g_{n}(t)$ is the instantaneous transitions out of state $|n\rangle$ for a jump to the other states $|j\rangle$ with specific rate $ k^{nc}_{jn}(t)$. Similarly the integral term account for the transitions that occur due to the past of the system.
}  

Classically in a semi-Markov process, 
for the survival probabilities functions to be meaningful, we must have 
\bea \label{gnns} 0\leq g_{n}(t) \leq 1,\,\dot g_{n}(t)\leq0, \forall t\geq0,\ena
where the first condition ensures that \( g_{n}(t) \) represents a valid probability, the second expresses that the likelihood of remaining in the current state decreases or remains constant over time: it follows  that the definition of the various $k_{nm}(t)$ is crucial for their validity. If the above conditions are fulfilled hence we are sure that both the populations and the transition functions are correct probabilistic functions. Hence $\rho_{nn}(t)\geq0, t\geq0$, which is, at least, a necessary conditions for the density operator to be positive. However, due to the form of the dynamical map, if we start from a special initial condition 
$
\rho(0)=\sum_{k}\alpha_{k} |k\rangle\langle k|$ with $\sum_n |\alpha_{k}|^2=1
$
then the above necessary conditions for a positive evolution are also sufficient. 
\vspace*{-0.3cm}\section{The epidemic model}\label{sect4}
In this model, we consider a specific application of the general framework presented earlier. We describe the epidemic dynamics by assuming that the population of infected individuals increases due to the virus, which acts on the reservoir of the healthy population.
Over time, infected individuals either recover or die. We assume that once individuals recover (or die), they cannot become infected again, making these processes fully irreversible. {This inherent irreversibility justifies using the GKSL equation with memory effects as a proper model for the system’s dynamics, since the GKSL framework naturally accounts for irreversible evolution. It is important, however, to distinguish between the origins of irreversibility in open quantum systems and in epidemic models. In open quantum systems, irreversibility arises primarily from taking a partial trace over the environmental degrees of freedom, leading to a loss of information about the environment, a feature captured by the GKSL formalism. In contrast, in epidemic models, the irreversibility is introduced as a phenomenological input that reflects biological realities, such as immunity or death, which prevent recovered individuals from being reinfected. By drawing an analogy where the healthy population serves as a reservoir and the various epidemic mechanisms are modeled via Lindblad operators, as described later  in this Section, we translate the mathematical structure of the GKSL equation into a framework that encapsulates the irreversible nature of biological processes.} 

To cast the system to obtain the evolution equations \eqref{eq1}-\eqref{eq3} we choose the following symbolic representation which helps us to simplify writing the formulae:
\be\label{41} \{|n\rangle\}_{n=0,\ldots 7}= \{|000\rangle, |001\rangle, |010\rangle, |011\rangle, |100\rangle, |101\rangle, |110\rangle, |111\rangle\}\en
where the order in the vectors follows the populations $\I$, $\R$, $\D$.
We make some basic and reasonable assumptions to derive the main operators inducing the dynamics for the populations $\I$, $\R$, $\D$.
Typically, the infectious process has no memory, with a peak after a certain number of days  $T_I$. We assume that this process is described by a time distribution $k_{T_I,\lambda_I}(t)$ having variance \(\lambda_I^2\). This suggests 
to introduce a first  Lindblad operators  that increases the mean value of $\I$ without acting on recovered or dead, so that most plausible choice is given by the  jump operator 
$\tilde L^{nc}_{\I}=|100\rangle \langle000|=|4\rangle\langle 0|.$ The action on the vector $|000\rangle$, i.e. to a state where there are no infected, deceased or recovered individuals, produce a new state of the system with infected individuals, but without (yet) dead or recovered.  Note that the creation operator \(\ha^\dagger_{\I}\) may be unsuitable in this context, as it will act also on the states \(|011\rangle, |010\rangle, |001\rangle\), thereby inducing an unintended infectious process among the recovered and dead populations, which contradicts the assumptions made. For instance $\ha^\dagger_{\I}|011\rangle=|111\rangle$.  The scenario of new infections among the recovered individuals will be considered separately later, together with the description of different waves of the epidemics.
Going back to $\tilde L^{nc}_{\I}$, we see that it only acts on the free-infection state $|000\rangle$ and projects it into the fully infected state $|100\rangle$. 

The recovery occurs, on average, after a certain number of days, denoted as \(T_R\), with the healing process following some distribution $k_{T_R,\lambda_R}(t)$ characterized also by a variance \(\lambda_R^2\).  Similarly, the mortality process has a mean time \(T_D\) with a time distribution $k_{T_D,\lambda_D}(t)$, with variance \(\lambda_D^2\). Both processes have memory because  they consider the status of the populations at previous time.
Hence the following  Lindblad operators must be considered:
$$\tilde L_{\R}=\ha^\dagger_{\R}a_{\I}=|2\rangle\langle 4| + |3\rangle\langle 5|,\qquad\tilde L_{\D}=\ha^\dagger_{\D}a_{\I}=|1\rangle\langle 4| + |3\rangle\langle 6|.$$
In view of the interpretation of the ladder operators we see that $\tilde L_{\R}$ is a {\it recovery operator}, while $\tilde L_{\D}$ is a {\it death operator}.
{Moreover, since no  reciprocal interactions occur among the populations $\I$, $\R$ and $\D$, and we want only to focus on the main and primary mechanisms ruling an epidemic event, the Hamiltonians in \eqref{eq2} and \eqref{eq3} are chosen to be zero, although one can define them different from zero in order to describe other mechanisms like fluctuations between recovered and infected not considered here.} {For instance,  epidemics processes can be characterized by more complex feedback reaction between the different states (e.g., a possibility of herd immunity development for some portions of the population, different reinfection rates,  and the severity of the actual infection). For such special cases, some phenomenological Hamiltonians can be introduced, to capture these interactive processes, taking place in the subsequent waves of epidemics.}

It follows that we can write the memory superoperator \eqref{eq2} as
\vspace*{-0.2cm}\bea
\K(t)[\bullet] = k_{T_D,\lambda_D}(t) \left( |1\rangle\langle 4| \bullet |4\rangle\langle 1| - \frac{1}{2} \{|4\rangle\langle 4|, \bullet \} +|3\rangle\langle 6| \bullet |6\rangle\langle 3| - \frac{1}{2} \{|6\rangle\langle 6|, \bullet \} \right)+\nonumber\\
k_{T_R,\lambda_R}(t) \left( |2\rangle\langle 4| \bullet |4\rangle\langle 2| - \frac{1}{2} \{|4\rangle\langle 4|, \bullet \} + |3\rangle\langle 5| \bullet |5\rangle\langle 3| - \frac{1}{2} \{|5\rangle\langle 5|, \bullet \} \right),
\ena
and the time local superopearator \eqref{eq3} as
\vspace*{-0.2cm}\bea
\Lc(t)[\bullet] = k_{T_I,\lambda_I}(t) \left(|4\rangle\langle 0| \bullet |0\rangle\langle 4| - \frac{1}{2} \{|0\rangle\langle 0|, \bullet \}
\right).
\ena
\vspace*{-0.0cm}It follows that the 
only non-zero functions \( k_{mn}(t) \) and  \( k^{nc}_{mn}(t) \) for superoperator \(\K\) and \(\Lc\) as in \eqref{eq2}-\eqref{eq3} are $
k_{24}(t) = k_{35}(t)=k_{T_D,\lambda_D}(t), \quad
k_{14}(t) = k_{36}(t)= k_{T_R,\lambda_R}(t), 	\quad	
k^{nc}_{40}(t) = k_{T_I,\lambda_I}(t).
$
Not surprisingly, a key aspect in modeling a given epidemic process is the proper definition of the functions $k_{T_I,\lambda_I}(t)$, $k_{T_R,\lambda_R}(t)$, and $k_{T_D,\lambda_D}(t)$. In this paper we assume the following general Gaussian behaviors for each function:
$$
k_{T_\sharp,\lambda_\sharp}(t) = \kappa_\sharp \exp\left(-\left(\frac{t - T_\sharp}{\lambda_\sharp}\right)^2\right),
$$
where $\sharp$ is any of $I,R,D$. In particular, the parameters $\kappa_I$, $\kappa_R$, and $\kappa_D$ determine the strength of the effects related to the respective Lindblad operators. In the case of $k_{T_I,\lambda_I}(t)$, the parameter $T_I$ sets the peak time of the infection process, which is closely related to the maximum number of infections per day, while $\lambda_I$ describes how sharply the infection process is peaked around $T_I$.
For the memory functions $k_{T_R,\lambda_R}(t)$ and $k_{T_D,\lambda_D}(t)$, the parameters  $T_R$ and $T_D$ represent the average time after which an infected individual recovers or dies, respectively. The parameters $\lambda_R$ and $\lambda_D$ measure the deviations, indicating the spread around the mean times $T_R$ and $T_D$.  

We can use $T_I$ as a characteristic time for the dynamics and define the new 
adimemnsional time as
$
t= \frac{t} {T_I}
$
so that the others parameters can be rescaled as
\be
T_R=\frac{T_R}{T_I},\, T_D=\frac{T_D}{T_I},\,\lambda_I= \frac{\lambda_I}{T_I},\,\lambda_R= \frac{\lambda_R}{T_I},\,\lambda_D= \frac{\lambda_D}{T_I},\,\kappa_I= T_I \kappa_I,\,\kappa_R= T_I \kappa_R,\,\kappa_D= T_I \kappa_D.
\en
\subsection{The initial condition and the equations for the populations}
The whole equations of motion for the populations, by appropriately reverting to the braket notation according to \eqref{41}, can be now written as follows\footnote{The equations are arranged in accordance with the order of the diagonal elements of the density matrix. {Here, for instance, $\rho_{|000\rangle}(t)$ is the (1,1) component of the $8\times8$ matrix $\rho(t)$, while $\rho_{|111\rangle}(t)$ is the  (8,8) component of the same matrix.}}:
\beano
\frac{d}{dt} \rho_{|000\rangle}(t) &=& -k_{T_I,\lambda_I}(t) \rho_{|000\rangle}(t) \\
\frac{d}{dt} \rho_{|001\rangle}(t) &=& \int_0^t d\tau \, k_{T_D,\lambda_D}(t - \tau) \rho_{|100\rangle}(t - \tau) \\
\frac{d}{dt} \rho_{|010\rangle}(t) &=& \int_0^t d\tau \, k_{T_R,\lambda_R}(t - \tau) \rho_{|100\rangle}(t) \\
\frac{d}{dt} \rho_{|011\rangle}(t) &=& \int_0^t d\tau \, \left[ k_{T_R,\lambda_R}(t - \tau) \rho_{|101\rangle}(t) + k_{T_D,\lambda_D}(t - \tau) \rho_{|110\rangle}(t) \right] \\
\frac{d}{dt} \rho_{|100\rangle}(t) &=&+ k_{T_I,\lambda_I}(t) \rho_{|000\rangle}(t) -\int_0^t d\tau \, \left[ k_{T_R,\lambda_R}(t - \tau) \rho_{|100\rangle}(t) + k_{T_D,\lambda_D}(t - \tau) \rho_{|100\rangle}(t) \right]  \\
\frac{d}{dt} \rho_{|101\rangle}(t) &=& -\int_0^t d\tau \, k_{T_R,\lambda_R}(t - \tau) \rho_{|101\rangle}(t) \\
\frac{d}{dt} \rho_{|110\rangle}(t) &=& -\int_0^t d\tau \, k_{T_D,\lambda_D}(t - \tau) \rho_{|110\rangle}(t) \\
\frac{d}{dt} \rho_{|111\rangle}(t) &=& 0
\enano
The interpretation of the various terms becomes straightforward when considering the assumptions made and the specific Lindblad operators proposed. To model the epidemic, we consider the initial state where all individuals are susceptible, represented by $\lvert 000 \rangle$, indicating that the system starts with no infected, recovered, or dead, and hence
the initial condition of the density matrix is given by $
\rho(0) = \lvert 000 \rangle \langle 000 \rvert.
$
{From a detailed view to the equations of motions above, we can check that, with this initial condition, only the states $\{|000\rangle,|001\rangle,|010\rangle,|100\rangle\}$ evolve, so that the equations of motions further simplify to}
\begin{eqnarray}
\frac{d}{dt} \rho_{|000\rangle}(t) &=& -k_{T_I,\lambda_I}(t) \rho_{|000\rangle}(t) \\
\frac{d}{dt} \rho_{|001\rangle}(t) &=& \int_0^t d\tau \, k_{T_D,\lambda_D}(t - \tau) \rho_{|100\rangle}(t) \\
\frac{d}{dt} \rho_{|010\rangle}(t) &=& \int_0^t d\tau \, k_{T_R,\lambda_R}(t - \tau) \rho_{|100\rangle}(t) \\
\frac{d}{dt} \rho_{|100\rangle}(t) &=&+ k_{T_I,\lambda_I}(t) \rho_{|000\rangle}(t) -\int_0^t d\tau \, \left[ k_{T_R,\lambda_R}(t - \tau) \rho_{|100\rangle}(t) + k_{T_D,\lambda_D}(t - \tau) \rho_{|100\rangle}(t) \right],\nonumber\\
\end{eqnarray}
{the other elements remaining zero for all times.}
The infection process can start, governed by the Lindblad operators that describe transitions such as from $\lvert 000 \rangle$ to $\lvert 100 \rangle$, representing the spread of infection, as well as subsequent transitions representing recovery and death.
This setup allows us to track the progression of the infection from the very beginning, in line with the assumption that the epidemic begins with no prior infections, recoveries, or deaths.
We highlight here that starting from the above initial condition, or any other given by a superposition of the states $|000\rangle,|001\rangle,|010\rangle,|100\rangle$, leads to 
\be\langle \I \rangle+ \langle \R \rangle+ \langle \D \rangle\leq1,\en
which allows for a phenomenological interpretation of the expected values as percentages of individuals belonging to the various compartments.

\section{Numerical Experiments}\label{sect5}
In this experiment, we aim to simulate the typical evolution of a seasonal flu outbreak (see real data in Italy, \cite{FLuIta}). We model realistic infection dynamics, where the number of infected individuals peaks after about 30 days, with recovery and mortality occurring approximately 3-4 days after infection. Therefore the scaled parameters are: $\kappa_I=0.07$, $\kappa_R=40$, $\kappa_D=6$, $\lambda_I=0.33$, $\lambda_R=0.035$, $\lambda_D=0.035$, $T_R=0.11$, $T_D=0.13$, with initial condition $\rho(0) = \lvert 000 \rangle \langle 000 \rvert$.

\begin{figure}[!h]	
\begin{center}
	\subfigure[ ]{\includegraphics[width=7.5cm]{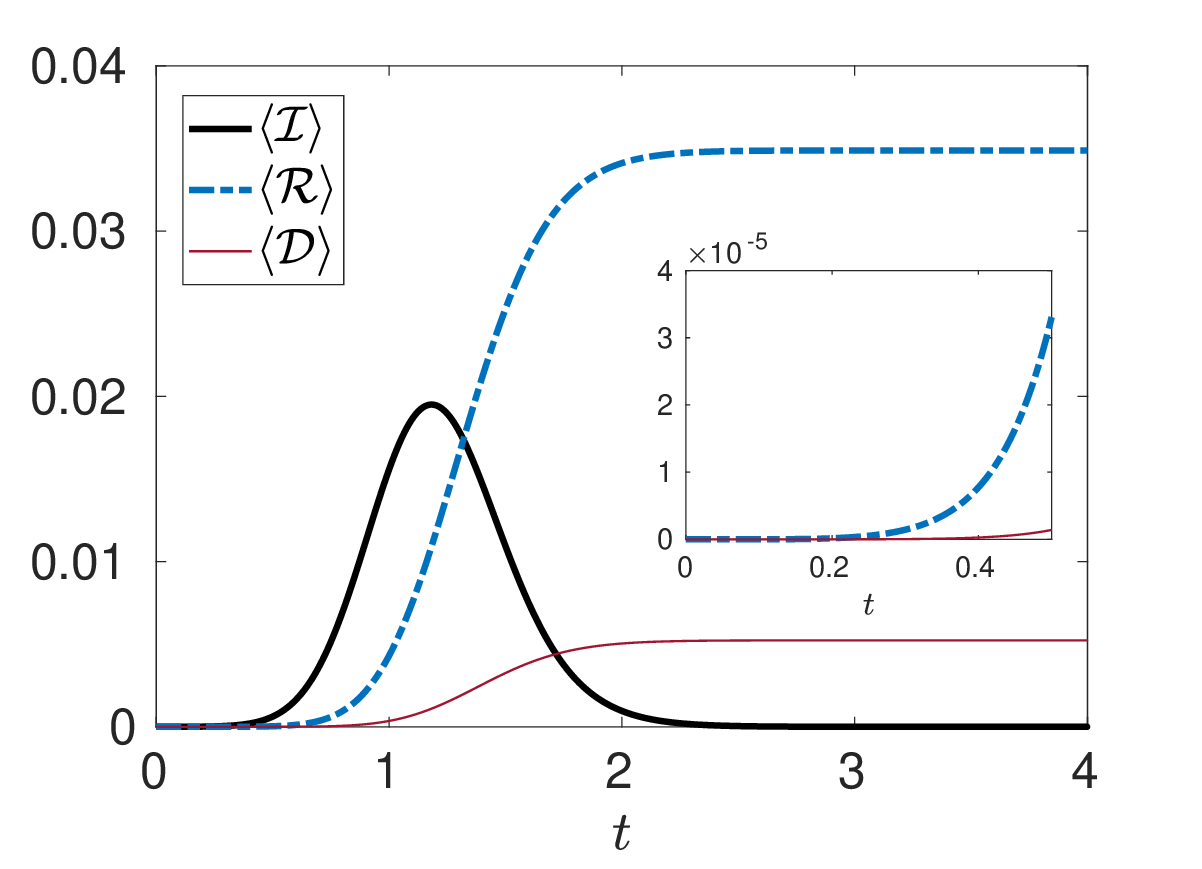}}	
	\subfigure[]{\includegraphics[width=7.5cm]{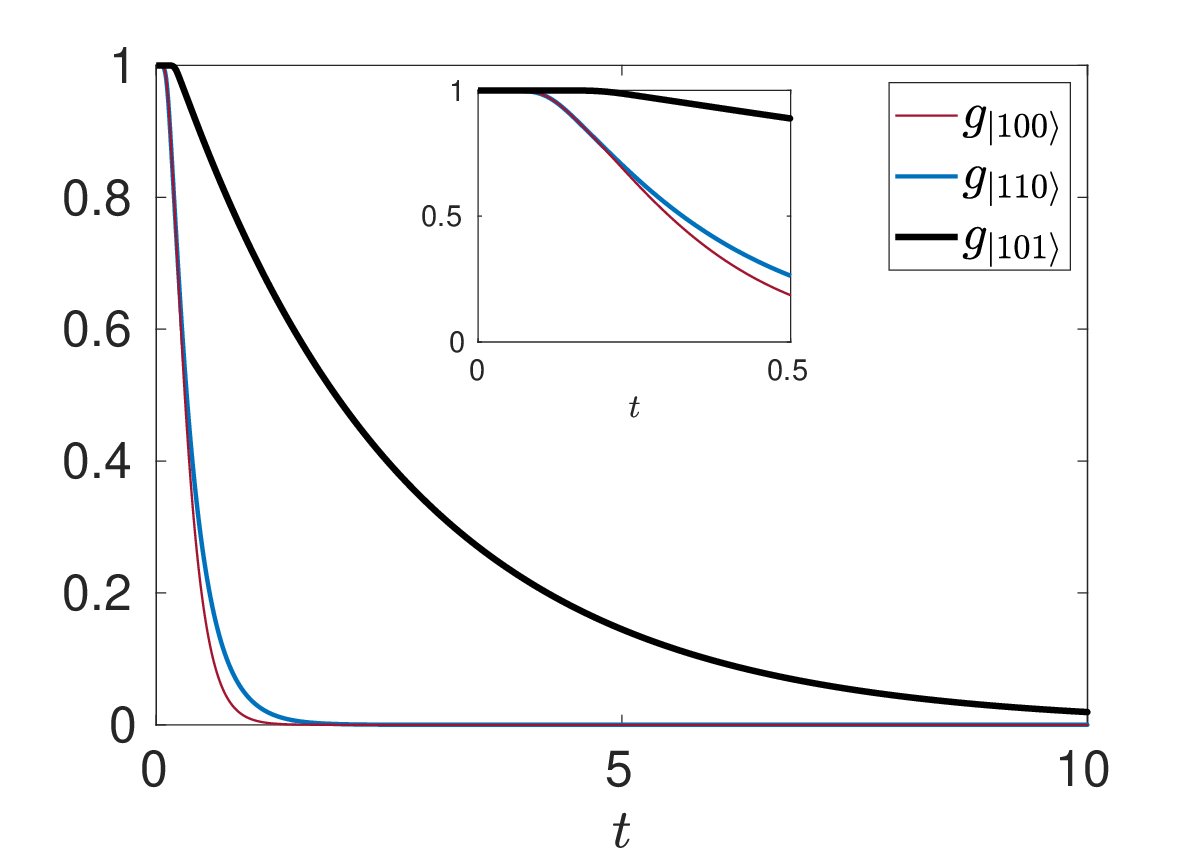}}\\
	\subfigure[]{\includegraphics[width=7.5cm]{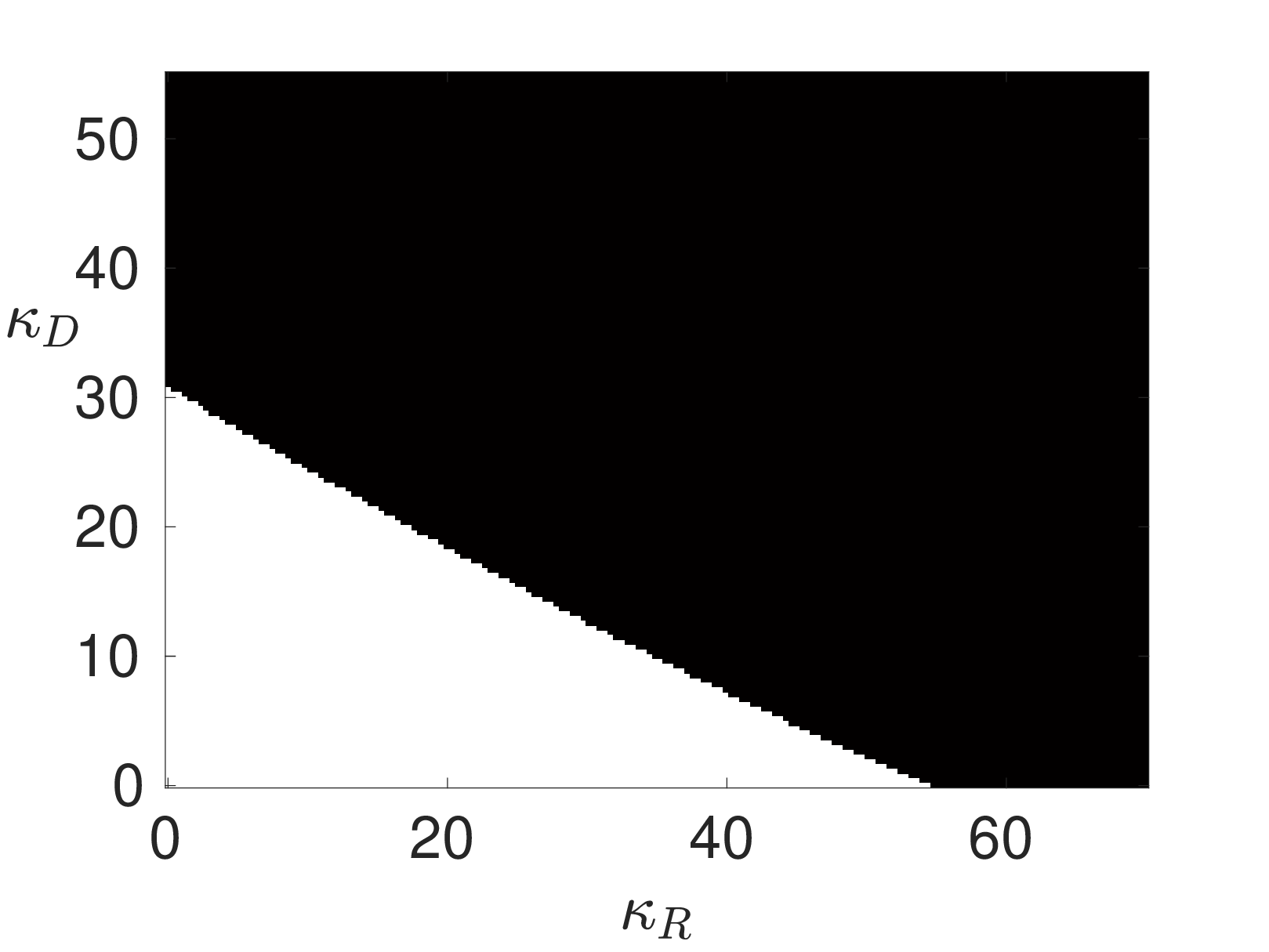}}
	\subfigure[ ]{\includegraphics[width=7.5cm]{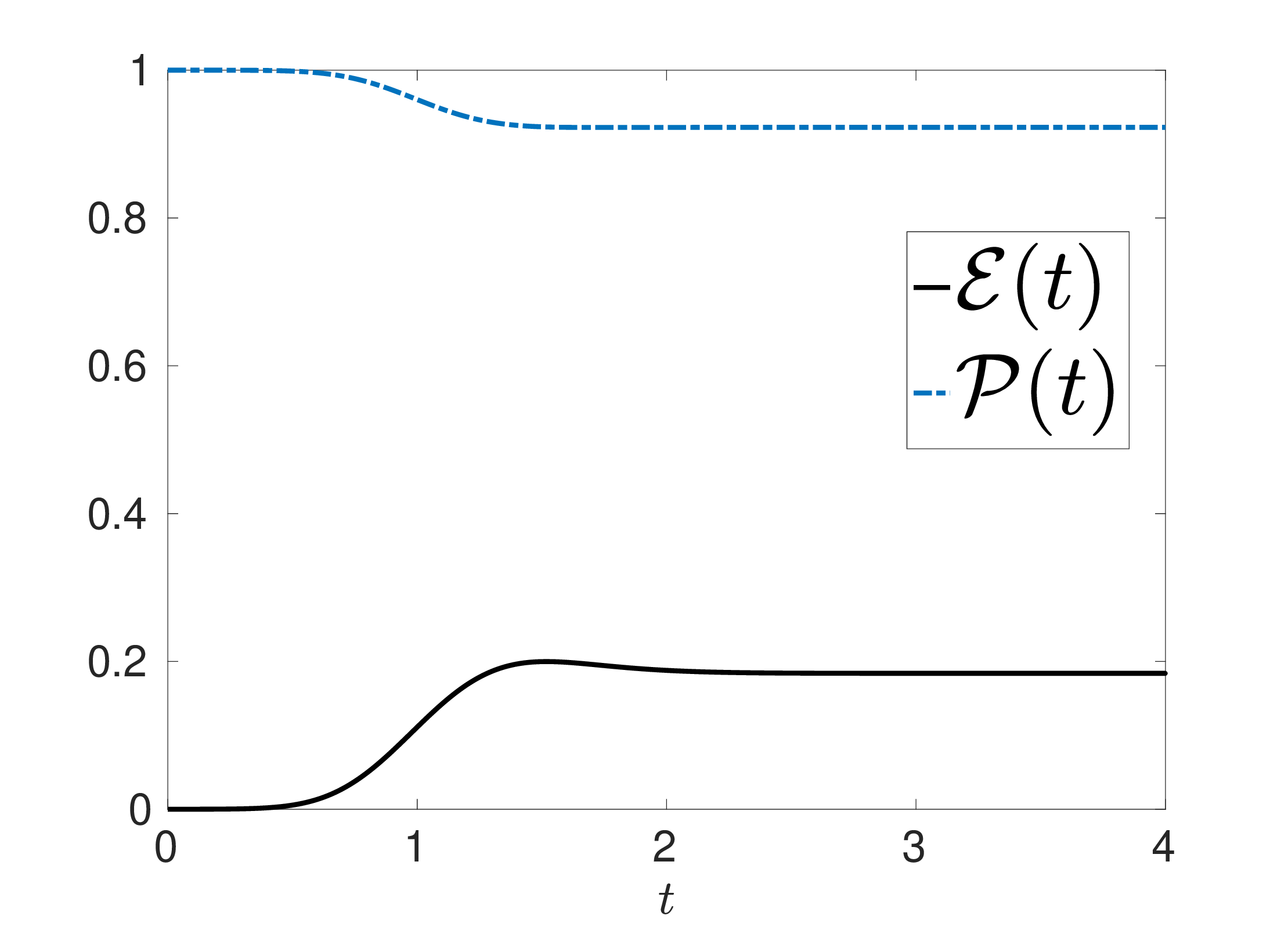}}	
	\caption{\textbf{(a)} Time evolutions of the expectation values $\langle \I \rangle,\langle \R \rangle,\langle \D \rangle$.\textbf{ (b)}
		Time evolutions of the survival probability functions $g_{|100\rangle}(t),g_{|101\rangle}(t),g_{|110\rangle}(t)$.
		\textbf{ (c)}
		Region of positivity-negativity  of the evolution for the survival probability function $g_{|100\rangle}$ in the  $\kappa_R-\kappa_D$ plane: white region admissible region, black region non admissible region. 	\textbf{ (d)} Time evolutions of the entropy and purity. }
	\label{fig:evo1}
\end{center}
\end{figure}

Figure \ref{fig:evo1}a shows the time evolution of $\langle \I \rangle$, $\langle \R \rangle,\langle \D \rangle$. The infected population increases, with the highest gradient peaking around $t \approx 1$ (30 days in dimensional time). {Recovered and dead populations increase later, as shown in the inset, and since $T_R > T_D$ and $\kappa_R > \kappa_D$, recovery is more probable and faster than mortality}. After $t \approx 1.18$, the infected population decreases, and recovered and dead populations reach their asymptotic values. At this point, the infection process concludes, with no further recoveries or deaths occurring. The equilibrium reflects characteristics of open quantum systems. We note that, while this is not the case here, equilibrium in a purely closed system can also be obtained using $(H,\rho)$-induced dynamics \cite{Gorgone2023, fffr}.
For this set of parameters, we checked the positivity of the evolution regardless of the initial condition, as shown in Figure \ref{fig:evo1}b. The evolution of the survival probability functions $g_{|100\rangle}(t)$, $g_{|101\rangle}(t)$, and $g_{|110\rangle}(t)$, which have non-trivial dynamics based on the values of $k_{nm}$, are plotted. All these functions remain positive for very large times and almost asymptotically decaying to zero. As expected, $g_{|100\rangle}(t)$ is most influenced by memory effects and decreases at the fastest rate.
As stated earlier, the functions $g_{n}(t)$ must satisfy conditions \eqref{gnns} to ensure a physically valid evolution consistent with the probabilistic nature of the process. The time evolutions in Figure \ref{fig:evo1}b confirm that positivity is preserved for this parameter set. However, altering the parameters governing memory effects could result in a loss of positivity. Since $g_{|100\rangle}(t)$ is more strongly affected by memory effects, we examined for which values of $\kappa_R$ and $\kappa_D$ positivity is maintained. This is shown in Figure \ref{fig:evo1}c, where the $\kappa_R,\kappa_D$ plane highlights the region of positivity: white indicates $g_{|100\rangle}(t) \geq 0$ for all $t \geq 0$, while black marks where $g_{|100\rangle}(t) \leq 0$ at least once. In the black region, the evolution is non-physical and inadmissible due to the appearance of negative probabilities.
\vspace*{-0.0cm}
We conclude by observing
that it is widely known that in the dynamics driven by the equations of an open quantum system, the stochastic nature of the quantum jumps induced by specific operators
induces a  mixture of states from a generic pure state that  increases the Von
Neumann entropy $\mathcal{E}$ and decreases the purity $\mathcal{P}$ defined as
\be
\mathcal{E}=-\text{Tr}(\rho \log \rho),\,\quad 
\mathcal{P}=\text{Tr}(\rho^2).
\en
Result are shown in Figure \ref{fig:evo1}d where it is evident the growth of the entropy and the loss of purity.
We highlight that this phenomenon starts from the beginning of the evolution, as can be seen using a standard perturbative approach for small time steps $dt$.
Without giving here further details, see for instances \cite{AsanoEtAl2013,BaGa2023}, the perturbed dynamics basically follows two distinct paths where on path is the discontinuous one characterized by quantum jumps to the new states (to be normalized in a standard way)
\bea
\tilde{L}^{nc}_\I \rho_0 \left.\tilde{L}^{nc}_\I\right.^\dagger,\,  \tilde{L}_\R\, \rho_0 \left.\tilde{L}_\R\right.^\dagger, \,  \tilde{L}_\D\, \rho_0 \left.\tilde{L}_\D\right.^\dagger,
\ena
These jumps are responsible for the typical mixture of states in dynamical systems governed by the equation of an open quantum systems, and very well captured by the growth in time of the entropy and by a parallel decrease of the purity.  %%% but we have stabilization and even a small fall in entropy (t=1.5), do you agree? If not, then the "camel-like" discussion is less relevant. 
{The relationship between entropy dynamics and epidemic dynamics can be explained by the transitions between compartments which contribute to state mixing and an increase in entropy. Each of these transitions introduces irreversible changes to the system, amplifying both the state mixture and the associated entropy growth. We can also interpret the dynamics of entropy—characterized by initial growth, followed by a slight decline and stabilization—in light of the adaptive behavior of living systems, as expounded in the work by \cite{Camel-like}. This study demonstrates that, under certain conditions, a biological system (e.g., the human body) adapts to specific environmental features, resulting in the emergence of "camel-like" humps in entropy behavior. In our specific example (cf. Figure \ref{fig:evo1}d), we observe that, following an initial period of adaptation to the epidemic environment, densities stabilize in the limiting states at time interval $\approx2$ days after the epidemics emergence. As shown in \cite{Camel-like}, under certain parameters of the Lindblad coefficients and the initial states of the infected individuals (biological systems), the growth and decline in entropy can become more pronounced, characterized by multiple "humps" and ultimately resulting in the restoration of coherence. This phenomenon is an interesting feature that may provide insight into human adaptation and recovery processes within a broader class of pandemics.\footnote{For biological systems interacting with an environment, non-unitarity has been identified as a necessary condition for inducing "camel-like" dynamics.}}

% can we say that $\tilde{L}_\I$ also encodes the specifics of internal characteristics i.e. the way it affects them (light/severe) and hence when the individuals transit to infected state the adaptation to the infection still takes place in their body (biological state) which is characterised by initial growth of entropy and a decrease of pureness. When this adaption took place (in real time 1.5 -2 months) the average individual has either adopted to the disease and recovered or his biological system did not manage to combat the disease and he/she transited to a deceased state. From graph \ref{fig:evo1}d we can observe that the entropy can even start falling (in our graph only slightly) under certain conditions. For instance we could also add that this shape of entropy and emergence of `humps' can manifest to various degrees when confronted with different infections (e.g. as captured in Lindblad operators and initial states of the individuals). 
% I have written it in more succinct way above. 
%My main questions is if the slight decrease in entropy in the graph \ref{fig:evo1}d is an artefact (at around t 1.5) or under certain conditions can be made more pronounced " camel -like". In any case we see that it definitely stabilises to an equilibrium state (as the individuals adopt to the infections environment, by either recovering or dying). 

{\section{A different application and conclusions}\label{sect6}

In this paper, we have utilized a GKSL equation with memory effects to describe the evolution of a generic (macroscopic) system. Specifically, we have examined an epidemic model in which infection occurs as an instantaneous process, whereas recovery and death of individuals occur only after a certain time interval.
This time delay has been modeled using a semi-Markovian effect, enabling an accurate description of the distinct nature of interactions occurring in the sub-system, $\Sc_P$.
{We should emphasize that we have not addressed here in details the issue of the complete positivity of the evolution, a well-known challenge in real physical systems governed by non-Markovian dynamics (see \cite{Budini,BreuerVacchini2008,Chru2021}). The reason is that our primary objective was to derive a set of equations exhibiting classical behavior, without accounting for typical quantum effects, such as entanglement and superposition, which are essential in real quantum physical systems and are indeed connected to the concept of complete positivity.} 
We stress however, that these quantum effects have been shown to have a role also  outside physical systems, for instance in in human cognition which does not adhere to the axiomatics of Boolean logic, leading to contextual human behavior in incompatible decision-making scenarios. For a historical account of the application of quantum probability to capture these features of human cognition, refer to the recent paper \cite{QuantumCognition} and the monographs \cite{Busemeyer2012} and \cite{havkhrebook}.
The presence of genuine contextuality in quantum cognition has also been confirmed through Bell tests in recent experimental studies, which highlight the emergence of contextuality from incompatible observables characterized by the indeterminacy of preferences (see \cite{Bruza2023}) and in the memory recollection of past events (see \cite{TempBell}).

We would like to conclude this paper with a comment on the potential applicability of the proposed model to a completely different problem in decision-making. Consider an agent, Alice, who, during a political election, needs to decide—based on information she receives from her environment—whether to vote for a right-wing or left-wing party and, in the same election, for a right-wing or left-wing candidate as president of the region (or province, or city) where she resides. %, or whether to abstain from voting altogether.
This scenario is typical of regional political elections where split-ticket voting is permitted: an elector may vote for a left-wing party while simultaneously preferring a right-wing candidate for the presidency of the region.\footnote{This phenomenon has been modeled using an open-systems framework for U.S. Presidential and Congressional elections, which are often characterized by voters' "ticket-splitting" behavior, see \cite{Elections2014}, \cite{Elections2016}. In that work, quantum Markovian dynamics was employed under the assumption that voters are essentially "memoryless," with the intensive informational "bath" erasing memories of older news and political events.} Alice's decision-making process requires time, during which she stores and processes information before reaching her final decision.
In this model,
$\langle \I \rangle$,
$\langle \R \rangle$, and
$\langle \D \rangle$
should be understood, respectively, as the total amount of political and related information reaching Alice, her time-dependent tendency to vote for a right- or left-wing party, and her (also time-dependent) tendency to vote for a right- or left-wing candidate. Naturally, these two decisions may fluctuate over time, but once the decision is made, their limiting probability distribution stabilizes. This behavior is illustrated, for instance, in Figure \ref{fig:evo1}(a), where the numbers on the axes do not have an actual interpretation but can be approximated with a real-time axis adapted to that setting. We can highlight the advancement of the presented semi-Markovian model in capturing the impact and duration of the political deliberation state, incorporating the effects of working memory and Alice’s initial state regarding her personal values and prior political affiliations.\footnote{Memory recollection, considering some major political dimensions (e.g., domestic policy, foreign policy, government affairs, etc.), and initial psychological attachment to specific parties and candidates is claimed to play a vital role in driving the final probability distribution of votes, as shown in the early foundational "funnel model" by \cite{Campbell}.}
In the devised model, what is important is the overall behavior of the three curves. The black line, representing the information reaching Alice, begins to increase from the very start of the time evolution (e.g., pre-election political campaign). As it increases, after some time, we observe that the two curves in blue and red begin to approach an asymptotic limit, which can be interpreted as Alice's final decisions. Hence, our conclusion is that Alice will most likely vote for, say, a right-wing party and a left-wing candidate. %based on limiting probability distributions 
We should also emphasize that the fact that the curve for information starts decreasing after a certain point models the situation in which Alice has already received all the information she is capable of processing, and no more information is needed. At this stage, she simply switches off the television and internet and processes the information with which she is satisfied.
This different point of view to the system in Section \ref{sect4} is just an indication that the explicit model considered in this paper can be adapted, with minor changes (as the explicit form of $k_{T_I,\lambda_I}(t)$, $k_{T_R,\lambda_R}(t)$, and $k_{T_D,\lambda_D}(t)$, to cite a possibility) to other interesting situations where interaction with an environment with some memory effect takes place, in variety different fields of research.% to other \\
\\

%\noindent \textbf{Acknowledgment}
%F.B. and F.G. acknowledge support under the National Recovery and Resilience Plan (NRRP) funded by the European Union - NextGenerationEU -
%	Project Title "Transport phenomena in low dimensional structures: models, simulations and theoretical aspects"- project code 2022TMW2PY - CUP B53D23009500006.	
%	F.B. and F.G. also  acknowledge support from the FFR2024 grant of the University of Palermo and support of the G.N.F.M. of the INdAM.  F.G. and P.K. acknowledge partial financial support under the "Networking" project of the Department of Engineering of the University of Palermo. F. B. was also supported by the PRIN-PNRR 2022, “CAESAR” - FAIR PE0000013, CUP J53C22003010006 and by project ICON-Q, Partenariato Esteso NQSTI - PE00000023, Spoke 2. P.K. was supported by Cost Action CA21169 "Information, Coding and Biological Function: the Dynamics of Life".
%

\end{document}